\documentclass[nofootinbib,secnumarabic,amssymb,nobibnotes,aps]{revtex4}
\usepackage{graphicx}
\usepackage{epstopdf}
\usepackage{subfigure}

\begin{document}

\rightline{\vbox{\halign{&#\hfil\cr
&MAN/HEP/2010/5\cr}}}
\rightline{\vbox{\halign{&#\hfil\cr
&IPPP/10/36\cr}}}
\rightline{\vbox{\halign{&#\hfil\cr
&DCPT/10/72\cr}}}

\title{Extracting Higgs boson couplings using a jet veto}
\author{B.E.~Cox}
\affiliation{School of Physics \& Astronomy, University of Manchester, Oxford Road. Manchester M13 9PL. U.K.}
\author{J.R.~Forshaw}
\affiliation{School of Physics \& Astronomy, University of Manchester, Oxford Road. Manchester M13 9PL. U.K.}
\author{A.D.~Pilkington}
\affiliation{Institute for Particle Physics Phenomenology, University of Durham, South Road. Durham DH1 3LE. U.K.}
\affiliation{School of Physics \& Astronomy, University of Manchester, Oxford Road. Manchester M13 9PL. U.K.}
\begin{abstract}
We show that the Higgs boson's effective couplings to gluons and to weak vector bosons can be extracted simultaneously from an analysis of Higgs plus dijet events by studying the dependence of the observed cross-section upon a third-jet veto.
\end{abstract}
\maketitle

\section{Introduction}

If electroweak symmetry is broken by the Higgs mechanism, it is very likely that one or more Higgs bosons will be detected at the LHC. Measurements of the spin, CP and couplings of the observed Higgs states will be required to determine the nature of the Higgs sector. In this letter, we propose to extract the Higgs boson's effective-couplings to weak vector bosons ($\Lambda _{\rm V}$) and to gluons ($\Lambda _{\rm g}$) by fitting the observed cross-section for $pp \to Hjj+X$ as a function of a central jet veto scale, $Q_0$. Specifically, we focus upon collisions in which a Higgs boson ($H$) is produced in between two hard jets ($j$) that are far apart in pseudo-rapidity. Our interest is in studying the behaviour of these events subject to the further constraint that there should be no third jet with transverse momentum above $Q_0$ in the pseudo-rapidity region between the two hard jets.

In the standard Higgs-plus-two-jet analyses, the central jet veto enhances the contribution of vector boson fusion (VBF) over gluon fusion (GF)  by virtue of the difference in $t$-channel colour flow. In particular, $Q_0$ is usually chosen to be as low as possible whilst remaining robust against uncertainties due to the underlying event (UE). As $Q_0$ is increased, the relative contribution of $gg \to H$ is enhanced. Therefore, if the cross-section is measured as a function of $Q_0$, then a fit of the form $\sigma(Q_0) = \Lambda_{\rm g} \sigma_{\rm g}^{\rm SM}(Q_0) + \Lambda_{\rm V}  \sigma_{\rm V}^{\rm SM}(Q_0)$ allows for the simultaneous extraction of the Higgs couplings to both vector bosons and to gluons provided the Standard Model (SM) predictions ($\sigma_{\rm g}^{\rm SM}(Q_0)$ and $\sigma_V^{\rm SM}(Q_0)$) are sufficiently well known.\footnote[1]{The interference between the two is negligible.}

To illustrate the point, we consider a 120~GeV Higgs boson decaying to $\tau^+\tau^-$, although the analysis should also be applicable to other decay channels and different Higgs masses. The parameters $\Lambda_{\rm g}$ and $\Lambda_{\rm V}$ measure deviations from the SM expectation and are sensitive to the following combination of Higgs boson decay widths: 

\begin{equation}
\Lambda_{\rm g} \sigma_{\rm g}^{\rm SM} \, \mathrm{BR}(H \to \tau \tau) \propto \frac{\Gamma_{\rm gg} \Gamma_{\rm \tau\tau}}{\Gamma_{\rm total}}~~{\mathrm{and}}~~\Lambda_{\rm V} \sigma_{\rm V}^{\rm SM}  \, \mathrm{BR}(H \to \tau \tau)\propto \frac{\Gamma_{\rm WW} \Gamma_{\rm \tau\tau}}{\Gamma_{\rm total}}~.
\end{equation}
The second relation assumes the ratio of $WWH$ and $ZZH$ couplings is the same as for the Standard Model, which is a generic expectation of custodially symmetric Higgs sectors. If new physics breaks this universality then $\Lambda_{\mathrm{V}}$ measures the effective coupling to weak vector bosons. 
By fitting to a linear combination of the SM cross-sections, we are assuming that any new physics generates deviations in the Higgs boson couplings but does not otherwise influence its production and decay.\footnote[2]{An example where this is not the case is the MSSM scenario where the Higgs boson is produced as a result of bottom quark fusion.} We note that the ratio $\Gamma_{\rm gg}/\Gamma_{\rm WW}$ may be obtained directly from such a measurement.

\section{The signal cross-section}

We use the Sherpa 1.2 Monte Carlo (MC) \cite{Gleisberg:2008ta} to generate samples of VBF and GF events in $pp$ collisions at $\surd s = 14$~TeV using the CTEQ6L parton distribution functions \cite{Pumplin:2002vw}. We generate $H+n$~parton matrix elements ($n =2,3$) with CKKW matching between the matrix elements and the parton shower \cite{Catani:2001cc}. We also invoke $K$-factors to ensure that the generated cross-sections, after matrix element plus parton shower matching, are equal 
to the corresponding NLO results of Campbell, Ellis \& Zanderighi after implementing their `weak boson fusion search cuts' \cite{Campbell:2006xx}. We find that $K_{\rm VBF} = 1.18$ and $K_{\rm GF}=2.11$. The latter re-scaling is comparable to the factor of 1.84 used by Andersen, Campbell and H\"oche \cite{Binoth:2010ra} to match Sherpa to MCFM \cite{Campbell:2010ff} at $\surd s = 10$~TeV. 

\begin{figure}
\centering
\mbox{ 
\subfigure[]{\includegraphics[width=.5\textwidth]{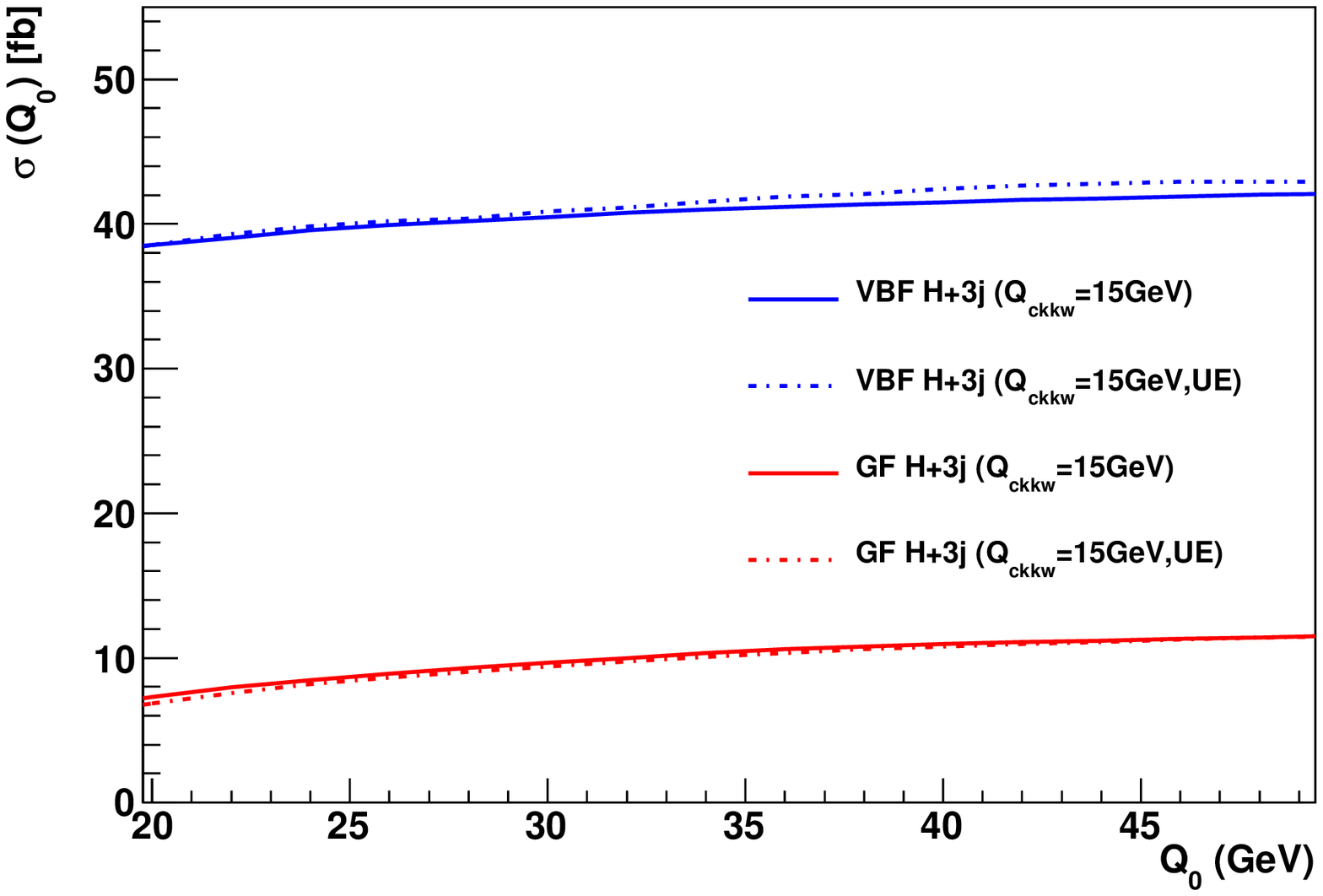}}
\quad       
\subfigure[]{\includegraphics[width=.5\textwidth]{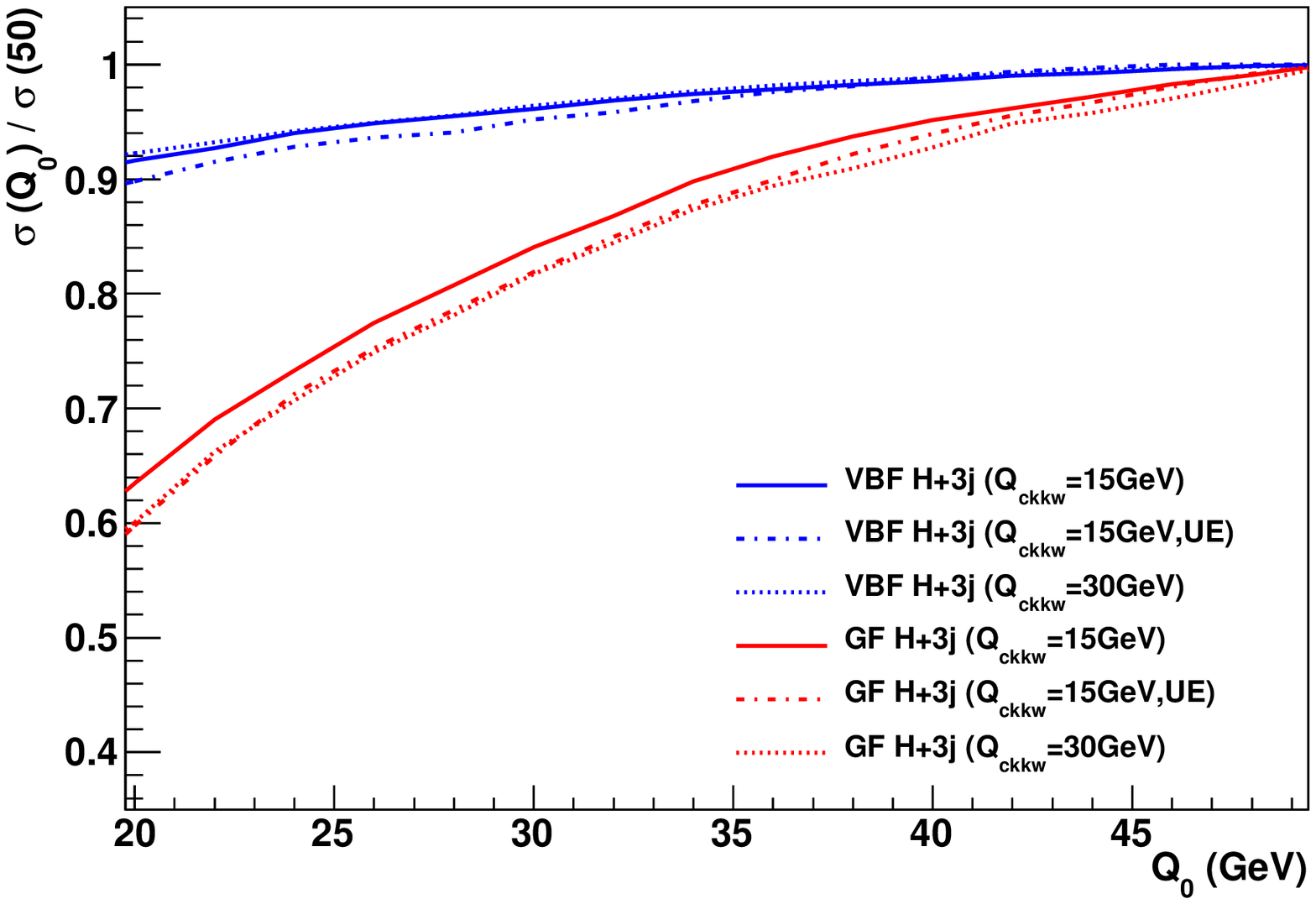}}
        }
\caption{The Vector Boson Fusion and Gluon Fusion contributions to the cross-section in the kinematic range described in the text for the process $pp \rightarrow H +jj+X$ with $H \to \tau \tau$ as a function of $Q_0$, the central jet veto scale: Figure (a) shows the absolute cross-section in fb; Figure (b) shows the cross-sections normalized to unity at $Q_0=50$~GeV. The default samples are labelled as $H+3j$ ($Q_{\rm ckkw}=15$~GeV). Also shown are the  effects of adding underlying event (UE) and changing the CKKW matching scale to $Q_{\rm ckkw}=30$~GeV. \label{fig:xs1}}
\end{figure}

In order to simulate the experimental acceptance, we apply the same cuts as those used in a recent ATLAS study \cite{atlas}. The anti-$k_{\rm T}$ algorithm is used with radius parameter $R=0.4$ to identify jets \cite{Cacciari:2008gp} and jets that fall within a cone of radius $\Delta R = 0.2$ around the true $\tau$ direction are identified as coming from a $\tau$-decay.\footnote{Experimentally, $\tau$'s are identified using dedicated algorithms and jets that overlap with these are identified in the same way.} The two $\tau$ candidates are required not to be back-to-back in azimuth, i.e. $\cos(\Delta \phi) > -0.9$, to facilitate the determination of the $\tau\tau$ invariant mass using the collinear approximation. Following \cite{atlas}, we apply further cuts on the highest transverse energy ($E_{\rm T}$) jets in the event excluding the $\tau$ jets:  
\begin{equation}\label{eq:cuts}
E_{\rm T,1} > 40~{\rm GeV,} \quad \qquad E_{\rm T,2} > 20~{\rm GeV,} \quad \qquad M_{\rm jj}>700~{\rm GeV,} \quad \qquad \Delta\eta>4.4, \quad \qquad \eta_1\times \eta_2 < 0~,
\end{equation}
where $\eta_{1,2}$ are the pseudo-rapidities of the jets, $\Delta \eta = |\eta_1 - \eta_2|$ and $M_{\rm jj}$ is the dijet invariant mass. 
In addition, the missing transverse energy is required to be greater than 30~GeV. 

Figure~\ref{fig:xs1}(a) shows the Vector Boson Fusion and Gluon Fusion contributions to the cross-sections (with the $K$-factors applied and after cuts) as a function of the central jet veto scale $Q_0$. Figure~\ref{fig:xs1}(b) shows the cross-sections as a function of $Q_0$ normalized to unity at $Q_0 = 50$~GeV in order to show the $Q_0$ dependence. Also shown are the effects of changing the CKKW matching scale from $15$~GeV to $30$~GeV and of adding the default underlying event in Sherpa. We place both of these uncertainties in context in the next section. 

In order to estimate the likely experimental event rates, the cross-sections shown in Figure \ref{fig:xs1} must be corrected to account for the experimental efficiencies $\epsilon_{\rm V}$ and $\epsilon_{\rm g}$ for selecting VBF and GF events respectively. These arise from the trigger efficiency, jet/$\tau$/lepton reconstruction and $\tau$/lepton identification. It is not possible to account for these effects in a study such as this, and we estimate $\epsilon_{\rm V} = 0.036$ for VBF using the numbers presented in \cite{atlas} for all $\tau$ decay channels. The efficiency also includes a veto on $b$-tagged leading jets to reduce the $t\bar{t}$ background. This is the only efficiency that is expected to vary between VBF and GF. However, this difference will be small relative to the overall normalization uncertainty on the GF cross-section and we therefore assume $\epsilon_{\rm g}=\epsilon_{\rm V}=\epsilon$.

We extract $\Lambda_{\rm g}$ and $\Lambda_{\rm V}$ using a pseudo-experiment approach. The number of events for each process (VBF or GF) is determined using a Poisson distribution. The mean is given by $\epsilon \Lambda_{i} \sigma_{ i}(50~{\rm GeV}) L$, where $L$ is the luminosity and $\sigma_i$ is the cross-section (shown in Figure~\ref{fig:xs1}(a)) at $Q_0=50$~GeV.\footnote{At this value of $Q_0$ we expect 91 VBF events and 25 GF events after all cuts with 60~fb$^{-1}$ of data.} Events are then selected at random from MC samples that survive the experimental cuts detailed above and a further cut on $Q_0$ allows the $Q_0$ distribution to be generated. Each point in the $Q_0$ distribution can then be smeared according to the expected systematic uncertainties, discussed below. The GF and VBF distributions are then combined to create the pseudo-data set. 

To extract the couplings, we fit the pseudo-data set to a function of the form $f (Q_0) = \Lambda_{\rm g}^{\prime} f_{\rm g} (Q_0) + \Lambda_{\rm V}^{\prime} f_{\rm V}(Q_0)$, where $f_{i}(Q_0) = \epsilon \, L \, \sigma_{i} \, (Q_0)$ are the SM theoretical predictions (i.e. the solid curves in Figure~\ref{fig:xs1}(a)). The whole procedure is repeated $10^4$ times for each value of $\Lambda_{\rm g}$ and $\Lambda_{\rm V}$ and the variation in the fitted values of $\Lambda_{\rm g}'$ and $\Lambda_{\rm V}'$ determines the accuracy to which we expect to measure $\Lambda_{\rm g}$ and $\Lambda_{\rm V}$. 

\begin{figure}
\centering
\mbox{       
\subfigure[]{\includegraphics[width=.35\textwidth]{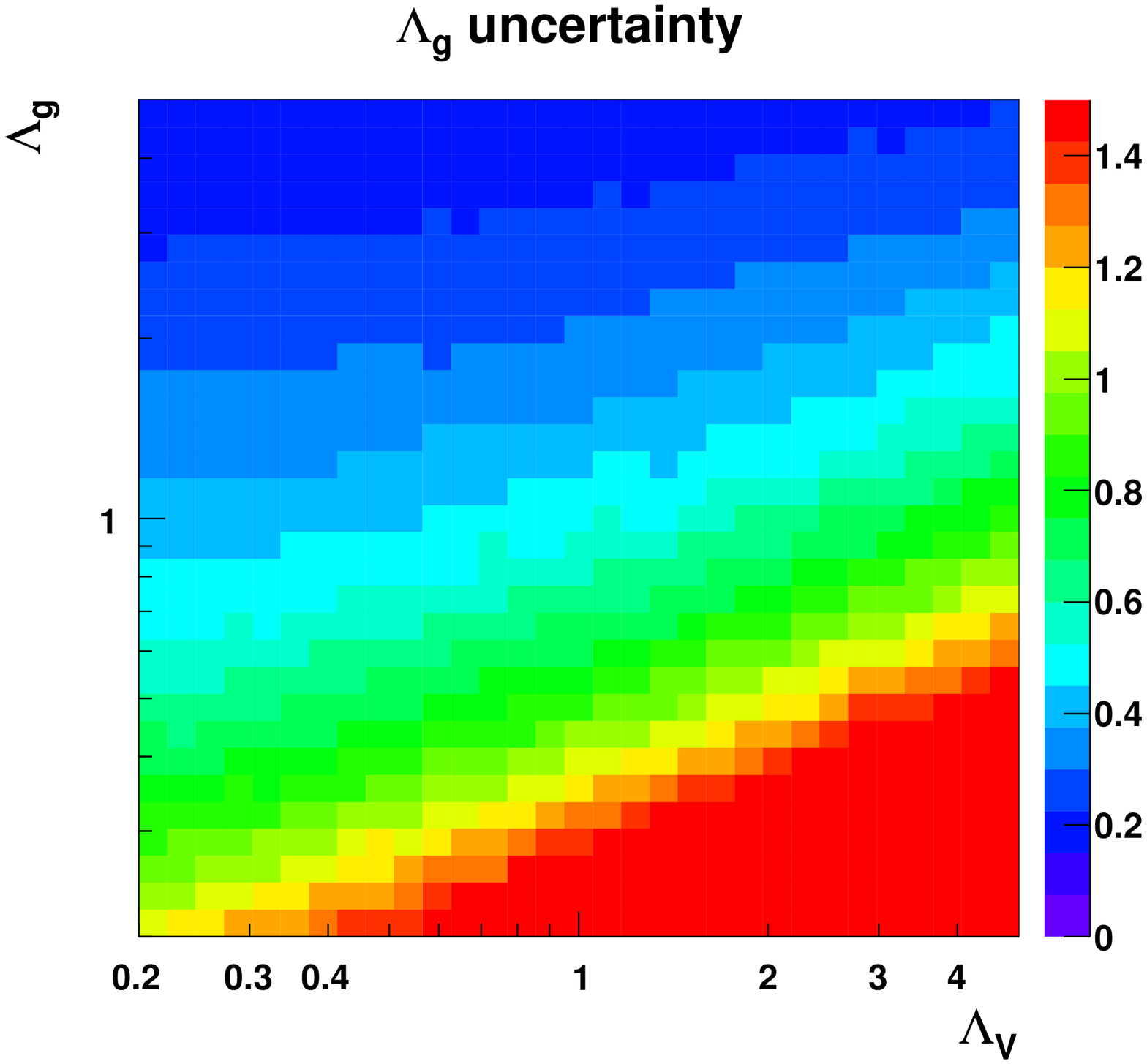}} \qquad \quad
\quad
\subfigure[]{\includegraphics[width=.35\textwidth]{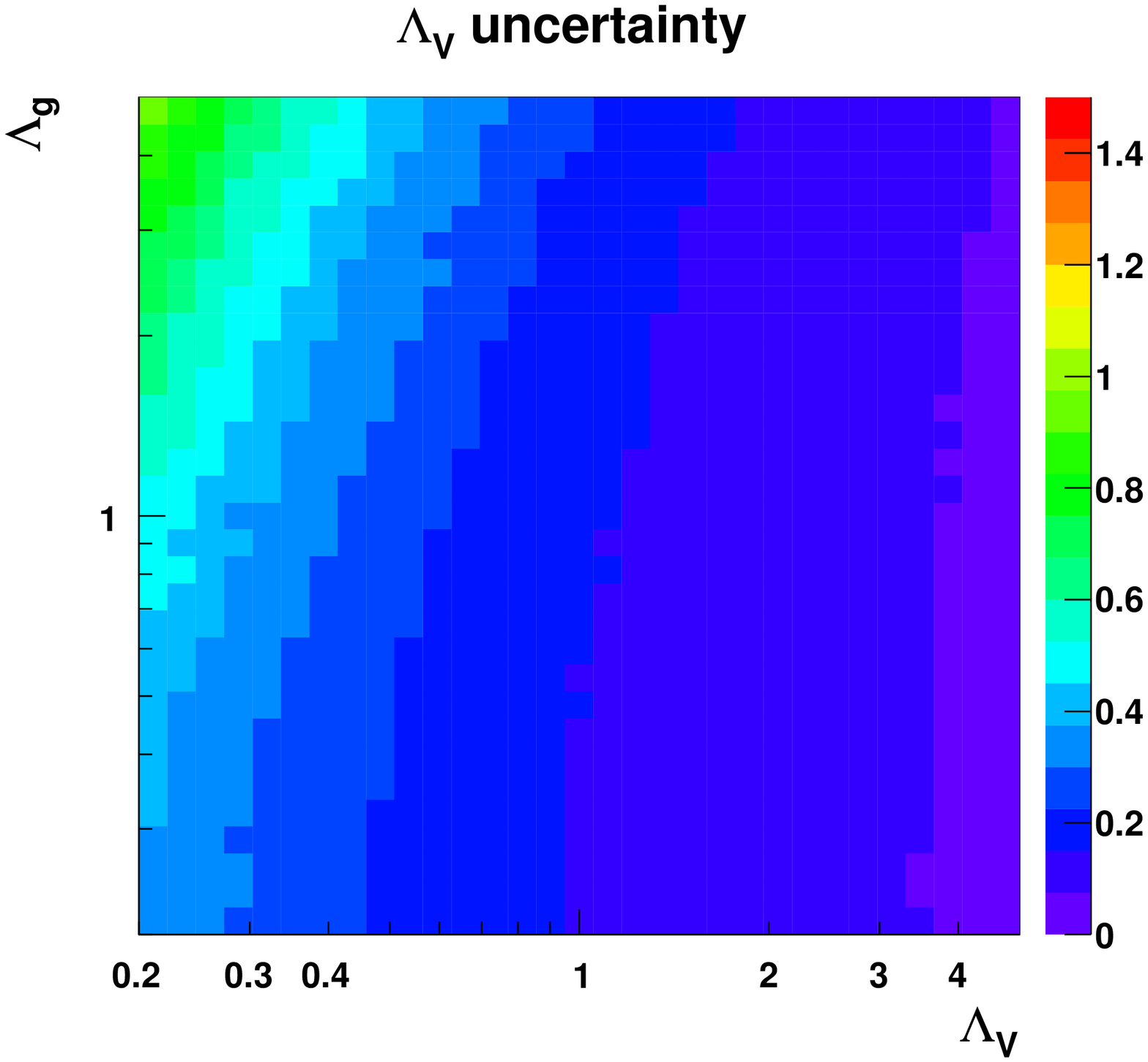}}
        }
\caption{The uncertainty in the extracted values of (a) $\Lambda_{\rm g}$ and (b) $\Lambda_{\rm V}$ given the default ATLAS cuts (see text) and 60 fb$^{-1}$ of data. Statistical uncertainties only. The scale on the right of each figure refers to the fractional uncertainty, i.e. $\delta \Lambda_{\mathrm{g,V}}/\Lambda_{\mathrm{g,V}}$.\label{fig:nosyst}}
\end{figure}

Figure \ref{fig:nosyst}(a) shows the expected uncertainty in the extracted value of $\Lambda_{\rm g}$ as a function of both $\Lambda_{\rm g}$ and $\Lambda_{\rm V}$ for $L=60$~fb$^{-1}$. The scale on the right of the figure refers to the fractional uncertainty, $\delta \Lambda_{\mathrm{g,V}}/\Lambda_{\mathrm{g,V}}$.
Figure \ref{fig:nosyst}(b) shows the corresponding uncertainty in $\Lambda_{\rm V}$. At this stage, we have ignored all theoretical uncertainties and experimental systematic errors. That $\Lambda_{\rm V}$ should be determined with greater accuracy over much of the parameter space is a result of the fact that the VBF cross-section is typically larger than the GF cross-section after cuts. 

\section{Effect of systematic uncertainties}\label{sec:theory}

The dependence of the cross-section on the central jet veto, $\sigma(Q_0)$, can be expressed as
\begin{equation}
\sigma(Q_0) = \sigma_{\rm jj}(1-P_{\mathrm{veto}}(Q_0))
\end{equation}
where  $\sigma_{\rm jj}$ is the $Hjj$ cross-section with no jet veto and $P_{\mathrm{veto}}(Q_0)$ is the probability of finding a third jet above $Q_0$ in the inter-jet region.
Uncertainties in $\sigma_{\rm jj}$ affect the overall normalization whilst uncertainties in $1-P_{\mathrm{veto}}(Q_0)$ also affect the shape of the $Q_0$ distribution. We can account for the theoretical uncertainty in the shape by estimating the uncertainties in  the predictions for $P_{{\rm veto}}$ at $Q_0=20$~GeV and at $Q_0=50$~GeV, which we treat as uncorrelated. In practice this means shifting the data points at these two extremal values of $Q_0$ in each pseudo-data set according to the theoretical uncertainties. For intermediate values of $Q_0$ we shift the pseudo-data points by interpolating linearly in $Q_0$ between the extremal shifts. 

Uncertainties in the VBF cross-section as a function of a third jet veto have been studied explicitly in the literature \cite{Figy:2007kv,Nason:2009ai}. In \cite{Figy:2007kv}, an estimate of the dominant NLO corrections to $Hjjj$ were studied, and it was concluded that $P_{\mathrm{veto}}(Q_0)$ is known to an absolute accuracy of better than $1\%$ for all relevant values of $Q_0$. In \cite{Nason:2009ai}, the NLO $Hjj$ cross-section was matched to the {\sc{herwig}} and {\sc{pythia}} parton showers in order to estimate $P_{\mathrm{veto}}(Q_0)$. The uncertainty was found to be larger ($\pm 3\%$) and comparable to the result of \cite{Figy:2007kv} for the LO $Hjjj$ calculation. We assume here that the yet-to-be-calculated NNLO $Hjj$ calculation with parton shower matching will confirm the NLO $Hjjj$ result of \cite{Figy:2007kv}. 

The overall normalization of the VBF cross-section is the other main theoretical uncertainty. The partial NNLO calculation presented in \cite{Bolzoni:2010xr} concludes that the overall uncertainty due to unknown higher-order QCD corrections is around $2\%$ (at $\surd s = 7$~TeV and without any VBF cuts). A complete NLO calculation, including electroweak corrections, is presented in \cite{Ciccolini:2007ec} where it is concluded that the total $Hjj$ cross-section with VBF cuts should be known to better than $2\%$ for Higgs masses in the range 100--200~GeV. On top of all these corrections, we must account for an overall uncertainty of some $3\%$ arising from uncertainties in the parton distribution functions \cite{Figy:2003nv,Berger:2004pca}. The net effect is that perturbative uncertainties on the VBF cross-section are probably no more than $\pm 4\%$ in the overall normalization and with no significant error on the $Q_0$ dependence. We implement this normalization uncertainty by varying all pseudo-data points by $\pm 4\%$, according to a flat distribution in $Q_0$. 

The theoretical uncertainties on the GF cross-section are considerably larger and the accuracy to which the $Q_0$ dependence of the cross-section can be computed is not well known.
The uncertainty in the NLO $Hjj$ calculation in the absence of a veto ($\sigma_{\rm jj}$) was estimated in \cite{Campbell:2006xx} to be around $\pm 20\%$ and we assume this value. We assign an additional, uncorrelated, uncertainty of $\pm 20\%$ to $1-P_{\mathrm{veto}}(Q_0)$ at $Q_0=20$ and $50$~GeV. The study in  \cite{Binoth:2010ra} suggests that the theory may be rather more uncertain at the present time \cite{Andersen:2008gc,Campbell:2006xx,Gleisberg:2008ta}. However, an uncertainty of the order that we assume here should be attainable in the near future (see also \cite{Forshaw:2007vb} for a study of the $Q_0$-dependence induced by wide-angle soft-gluon emission). The uncertainty due to CKKW matching illustrated in Figure \ref{fig:xs1} is clearly small on the scale of these other uncertainties.

Perturbative uncertainties are not the entire story. Chief amongst other theory uncertainties is the lack of knowledge of the underlying event (UE). In Figure~\ref{fig:xs1},  we illustrate the potential impact of the UE. The blue dot-dashed curve shows our prediction after including Sherpa's simulation of the UE and the blue solid curve is that in the absence of any UE. We assume an uncorrelated uncertainty of $\pm 1\%$ in $1-P_{\mathrm{veto}}(Q_0)$ at both $Q_0=20$~GeV and $50$~GeV in the VBF cross-section. In the case of GF, the effect of the UE is increased and we assign an uncorrelated $\pm 3\%$ uncertainty at $Q_0=20$~GeV and $50$~GeV. Early LHC data will allow tuning of UE models which should reduce the degree of uncertainty.

In addition to theoretical uncertainties we need also to estimate the impact of experimental systematics. The dominant systematic is due to the uncertainty in the jet energy scale (JES). The ATLAS collaboration estimated the total systematic uncertainty on the VBF cross-section to be $\pm 20\%$ at $Q_0=20$~GeV \cite{atlas}, of which 16\% is due to the JES and 10\% to other sources. We can satisfactorily reproduce the uncertainty arising from the JES by shifting the generator level jet energies by the ATLAS defaults (i.e. 7\% for $|\eta|<3.2$, 15\% otherwise). Importantly, we observe no significant dependence of the uncertainty from the JES on $Q_0$ and therefore assign an overall normalization uncertainty to the VBF cross-section of $\pm 20\%$. The uncertainty from the JES is larger for GF events, due to the steeper leading jet $E_{\rm T}$ and $M_{\rm jj}$ distributions, and it has a mild $Q_0$ dependence. We therefore estimate an overall normalization uncertainty of $\pm 30\%$ in conjunction with an additional uncorrelated uncertainty of $\pm 3\%$ in $1-P_{\mathrm{veto}}(Q_0)$ at $Q_0=20$~GeV and $50$~GeV. We also assume that the effects from the pile-up of multiple proton-proton interactions will be well understood and do not assign an additional uncertainty from this source.

To account for the influence of background processes, we assume that statistical fluctuations dominate.\footnote{The backgrounds are discussed in detail in \cite{atlas}: $Zjj$ is the most important background but $t\bar{t}$ production, $W+$jets production and QCD multi-jet production are also relevant.}   This should be the case if the background can be extracted by fitting the $m_{\tau\tau}$ distribution in data. We assume equal numbers of signal and background events at $Q_0=50$~GeV after cuts, which is broadly in line with the conclusions in \cite{atlas}. We also assume that the background varies with $Q_0$ such as to lie halfway between the upper and lower curves in Figure \ref{fig:xs1}(b). Our treatment of the background is clearly very approximate but is 
sufficient for the purpose of generating additional statistical fluctuations across the $Q_0$ distribution. We
note that the $Q_0$ dependence of the background will be determined from data.

\begin{figure}
\centering
\mbox{       
\subfigure[]{\includegraphics[width=.35\textwidth]{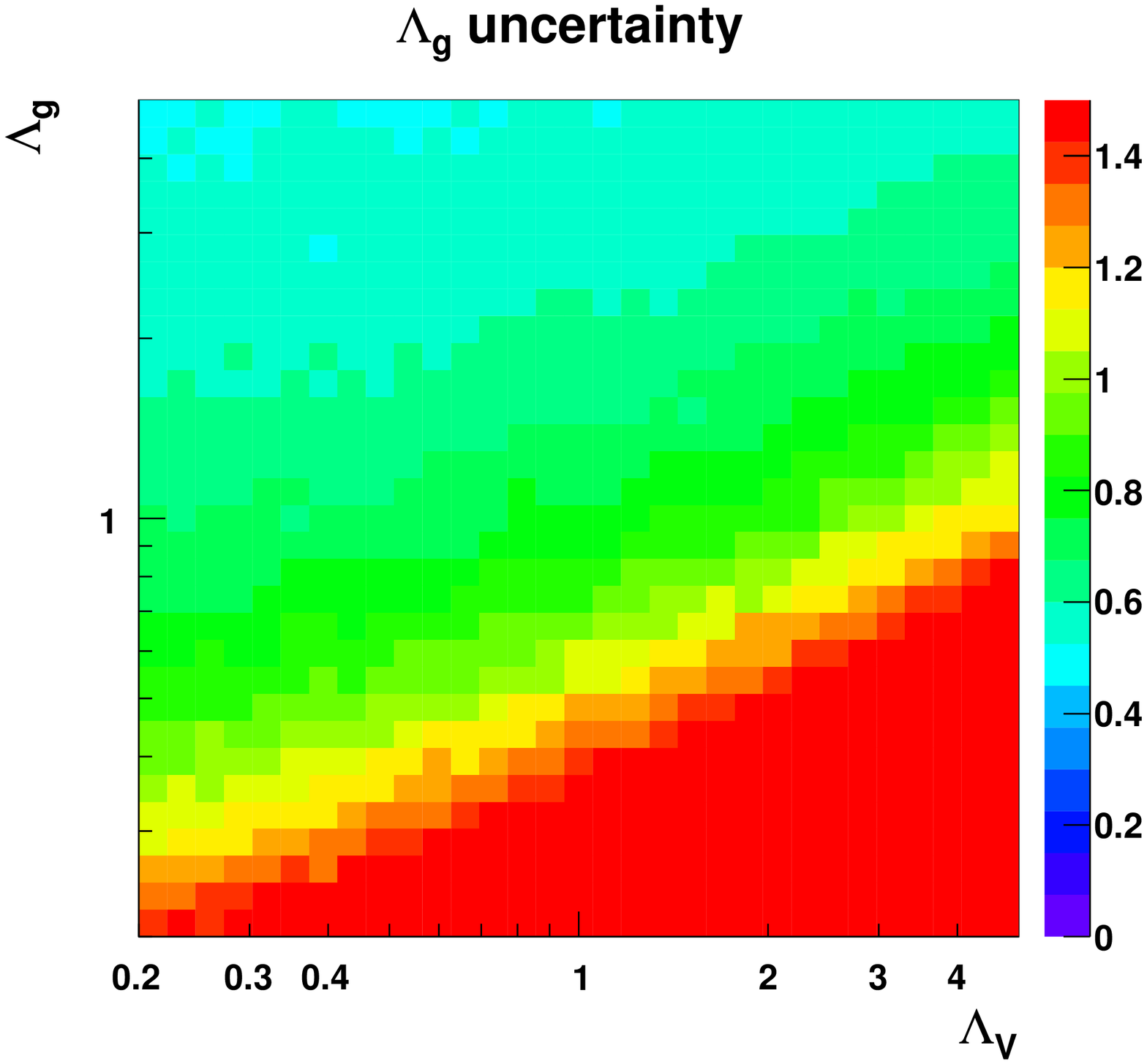}}\qquad \quad
\quad   
\subfigure[]{\includegraphics[width=.35\textwidth]{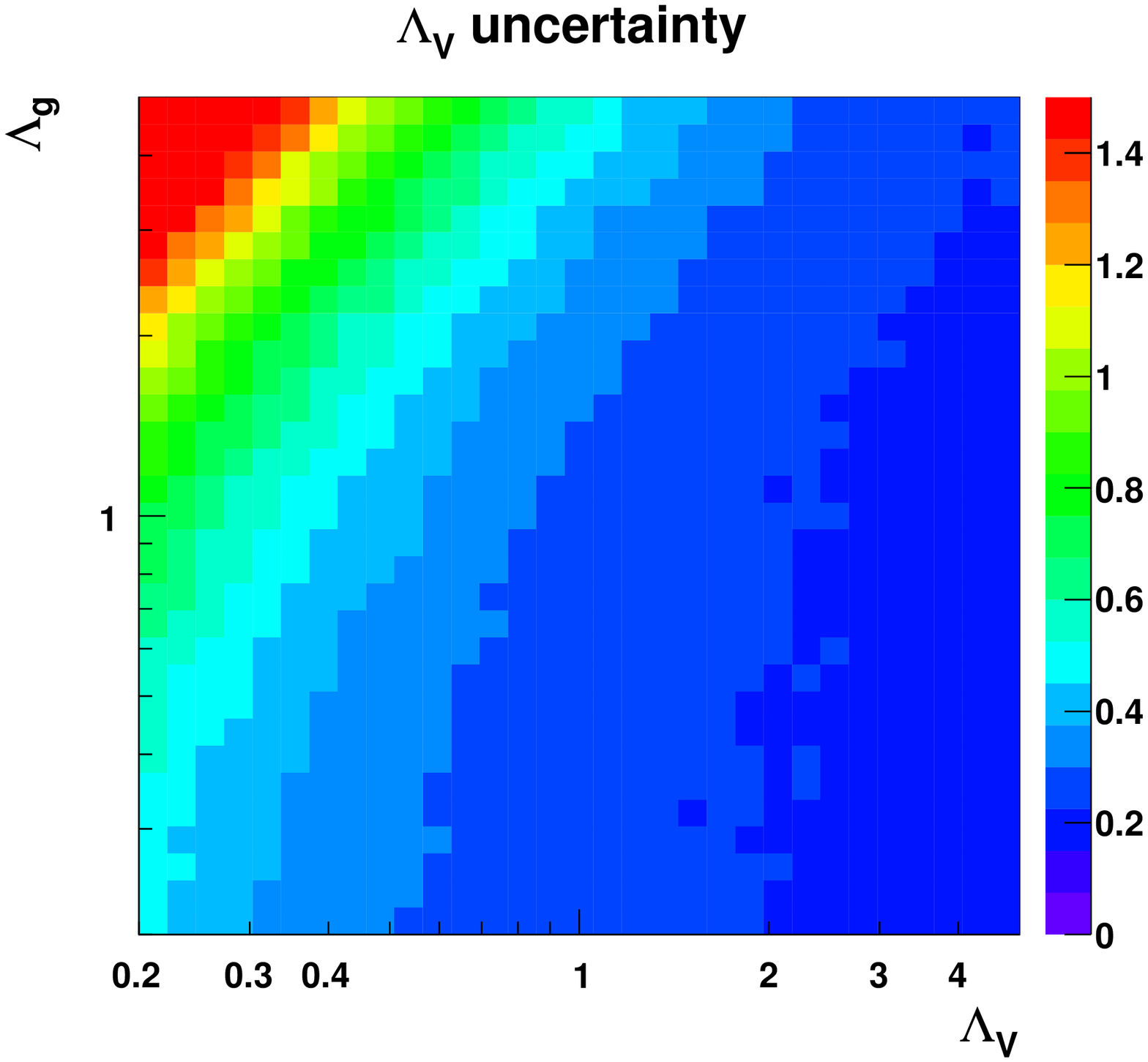}}
        }
\caption{The uncertainty in the extracted values of (a) $\Lambda_{\rm g}$ and (b) $\Lambda_{\rm V}$ given the default ATLAS cuts (see text) and 60 fb$^{-1}$ of data. The effects of systematic errors are estimated as discussed in the text. The scale on the right of each figure refers to the fractional uncertainty, i.e. $\delta \Lambda_{\mathrm{g,V}}/\Lambda_{\mathrm{g,V}}$.\label{fig:syst}}
\end{figure}

We are now in a position to quantify the effect of these various systematic uncertainties. Table \ref{tab:syst} illustrates the effect of the individual experimental and theoretical uncertainties on the extraction of the effective couplings and Figure \ref{fig:syst} shows the expected uncertainty after the inclusion of all systematic uncertainties. In the table, the numbers in the first row correspond to ignoring all experimental and theoretical uncertainties except those due to statistics and those in the final row represent the uncertainties after accounting for the various uncertainties discussed above. The middle four rows show the effect of including only the uncertainties on (i) statistical fluctuations in the background;  (ii) the VBF cross-section; (iii) the GF cross-section; (iv) experimental systematics. We also show results for a statistically comparable `BSM' point (the couplings are chosen so that the cross-section is approximately equal to that in the SM at $Q_0=50$~GeV).

\begin{table}[htdp]
\begin{center}
\begin{tabular}{|l||c|c||c|c|}
\hline
& \multicolumn{2}{|c||}{SM ($\Lambda_{\rm{g,V}}=1$)} & 
\multicolumn{2}{|c|}{BSM ($\Lambda_{\rm{g}}=4$, $\Lambda_{\rm{V}}=1/4$)} \\
 Error & $\sigma_{\Lambda_{\rm g}}/\Lambda_{\rm g}$ & 
$\sigma_{\Lambda_{\rm V}}/\Lambda_{\rm V}$ & $\sigma_{\Lambda_{\rm 
g}}/\Lambda_{\rm g}$ & $\sigma_{\Lambda_{\rm V}}/\Lambda_{\rm V}$  \\
\hline \hline
Stat.~only & 0.51 [0.23]& 0.16 [0.07] & 0.19 [0.08] & 0.72 [0.33] \\
\hline \hline
Backgd. & 0.56 [0.25] & 0.18 [0.08] & 0.20 [0.09] & 0.79 [0.35] \\
VBF & 0.52 [0.25] & 0.17 [0.08] & 0.19 [0.08] & 0.75 [0.33] \\
GF & 0.65 [0.45] & 0.19 [0.11] & 0.43 [0.40] & 1.56 [1.40] \\
Expt. & 0.62 [0.39] & 0.26 [0.21] & 0.35 [0.31] & 0.89 [0.52] \\
\hline \hline
All & 0.77 [0.57] & 0.28 [0.23] & 0.53 [0.50] & 1.66 [1.49] \\
\hline
\end{tabular}
\end{center}
\caption{Fractional error in extraction of $\Lambda_{\rm g}$ and 
$\Lambda_{\rm V}$ for the SM and a statistically comparable BSM 
parameter point. Numbers correspond to 60~fb$^{-1}$ of data or, in 
square brackets, to 300~fb$^{-1}$ of data.
\label{tab:syst}}
\end{table}%

\section{Conclusions}
The method we propose permits the measurement of the ratio of the effective couplings of the Higgs boson to gluons and to weak vector bosons within a single analysis. If the branching ratio to $\tau$ leptons is known, then the method can be used to extract the individual couplings. The couplings can be determined to an accuracy that is comparable to other methods \cite{Duhrssen:2004cv,Lafaye:2009vr}. A model-independent measurement of the relative size of the VBF and GF contributions should be of considerable value in subsequent analyses to study the CP nature of the $HWW$ \cite{Plehn:2001nj} and $Htt$   \cite{Klamke:2007cu,Andersen:2010zx} couplings via the azimuthal angle dependence of the tag-jets.

Our results demonstrate the need for an improved theoretical understanding of the veto dependence of the gluon fusion contribution. The primary experimental uncertainty comes from the knowledge of the jet energy scale.

We have not made any attempt to optimize the experimental cuts for this analysis, preferring instead to follow the study presented in \cite{atlas}, which was optimized for VBF production. It should therefore be possible to improve on our estimates of the coupling to gluons. Our method could be further generalized to include other $Hjj$ production channels, e.g. $b\bar{b}\to H$ in the case of MSSM scenarios in which the Higgs has an enhanced coupling to $b$-quarks.   

Using early LHC data, important precursors to the analysis presented here would be to perform analyses of the jet-veto dependence in dijet \cite{Forshaw:2009fz}, $W/Z+j$ and $W/Z+jj$ events \cite{Rainwater:1996ud}. 

\section{Acknowledgements}
We thank Jeppe Andersen, Keith Ellis, Frank Krauss, Mark Owen, Tilman Plehn, Frank Seigert, Soshi Tsuno and Mike Seymour for many very helpful discussions and the UK's Royal Society and STFC for financial support.

\bibliography{couplings}

\end{document}